\documentstyle[11pt,moriond,epsfig]{article}

\def\Journal#1#2#3#4{{#1} {\bf #2}, #3 (#4)}

\def\PRD{{\em Phys. Rev.} D}

\def\be{\begin{equation}}
\def\ee{\end{equation}}
\def\bea{\begin{eqnarray}}
\def\eea{\end{eqnarray}}

\begin{document}
\vspace*{4cm}
\title{PROBING $c\to u\gamma$ IN $B_c\to B_u^*\gamma$ DECAY 
\footnote{Talk  presented by Sa\v sa Prelov\v sek at XXXIV Recontres de Moriond on 'Electroweak interactions and unified theories', Les Arcs, 13-20.3.1999. Transparencies 
available at WWW pages at {\sf http://www.moriond.in2p3.fr/EW/transparencies}.} 
}

\author{ S. PRELOV\v SEK$^{a}$, S. FAJFER$^{a,b}$ and P. SINGER$^{c}$ }

\address{a) J. Stefan Institute, Jamova 39, 1001 Ljubljana, Slovenia \\
b) Department of Physics, University of Ljubljana, 1000 Ljubljana, 
Slovenia\\
c) Department of Physics, Technion - Israel Institute  of Technology, 
Haifa 32000, Israel}

\maketitle\abstracts{Flavour changing neutral  
 current (FCNC)
transitions $c\to u\gamma$ and $c\to ul^+l^-$ are very rare in the standard 
model and present an interesting probe for the physics beyond it. We study long 
distance contamination to these short distance processes in different hadron 
decays.  As the most suitable probe for $c\to u\gamma$ transition we propose 
$B_c\to B_u^*\gamma$ decay. Its detection at the branching ratio well above 
$10^{-8}$ would 
signal new physics. The $c\to ul^+l^-$ transition may be probed at high $m_{ll}$ 
in $D\to \pi l^+l^-$. The decays $D\to V\gamma$ and $D\to Vl^+l^-$ ($V$ is a 
light vector meson) are less 
suitable, since they are 
dominated by the long distance effects. }
 
\section{Introduction}

Flavour changing neutral current (FCNC) transitions occur in the standard model only at 
the loop level, hence they are very rare in the standard model and present 
a suitable probe for new physics. The rate for the FCNC transition strongly 
depends on the flavour of the quarks running in the loop: quarks of higher 
masses give higher rates, as long as their mixing with the external quarks is 
not highly suppressed.  From this aspect the FCNC transitions among down-like 
quarks $d$, $s$ and $b$ are relatively frequent: $s\to d$ transitions have been 
widely studied and recently also the $b\to s$ transition, enhanced due to the 
large top mass, has been observed by CLEO and ALEPH \cite{PDG}. 

The FCNC transitions among the up-like quarks $u$, $c$ and $t$  are especially 
rare in the standard model due to the small masses of the intermediate 
down-like quarks and only the upper experimental limits for this processes are 
available at present \cite{PDG}. A possible new heavy state in the loop could 
greatly enhance the rate of these transitions and they represent a suitable 
probe for new physics with almost zero background from the standard model. In 
present paper we explore the $c\to u\gamma$ and $c\to ul^+l^-$ decays, which 
are the most probable FCNC decays among the up-like quarks in the standard 
model.

In practice, the quark decays are probed in hadron decays.  A given
hadron decay is induced by the corresponding quark decay at short distances 
(SD), but it is  contaminated by the more mundane effects of
long distance (LD) dynamics such as the ones induced by propagating
intermediate hadrons. We examine SD and LD 
contributions to different hadron decays within the standard model, proposing 
the least LD contaminated ones as probes for  $c\to u\gamma$ or $c\to ul^+l^-$  
transitions or possible new physics. The most suitable probe for $c\to u\gamma$ 
transition
is found to be $B_c\to B_u^*\gamma$ decay \cite{bc}; the $B_c$ meson has been 
detected recently by CDF \cite{CDF}. Probing $c\to u\l^+\l^-$ might be possible 
at high $m_{ll}$ in $D\to \pi l^+l^-$ decay \cite{pll}  or at low $m_{ll}$ in 
$B_c\to B_ul^+l^-$ decay. 

\section{$c\to u\gamma$ and $c\to ul^+l^-$ at short distances}

The $c\to u\gamma$ is induced by the penguin diagrams with $d$, $s$
and $b$ quarks running in the loop, the dominant contribution coming from 
intermediate $d$ and $s$ quarks (in spite of large $m_b$ the $b$ quark 
contribution is suppressed due to small $V^*_{cb}V_{ub}$). The amplitude is 
strongly GIM suppressed at one loop giving the rise
to only $BR(c\to u\gamma)\sim 10^{-18}$. Including the
QCD corrections at the leading log approximation, the responsible Willson
coefficient $c_7$ (suppressed at one loop) obtains the admixture
of the other Willson coefficients (not all of them are suppressed at
one loop) 
and the amplitude is enhanced by two orders of magnitude
\cite{GHMW}. The complete two-loop QCD corrections further increase the 
amplitude by three orders and at this stage the $c\to u\gamma$ is induced by 
\cite{GHMW}
\begin{equation}
{\cal L}_{SD}^{c\to u\gamma}=-{G_F\over \sqrt{2}}{e\over 
8\pi^2}V_{cs}V_{us}^*c_7(\mu)\bar 
u\sigma^{\mu\nu}[m_c(1+\gamma_5)+m_u(1-\gamma_5)]cF_{\mu\nu}~,\qquad 
c_7(m_c)=-0.0068-0.020i
\end{equation} 
 giving the rise to
$\Gamma(c\to u\gamma)/\Gamma(D^0)\sim 2.5\times 10^{-8}$. Including further
QCD corrections  further increase of the rate is not expected
\cite{GHMW}.  

The $c\to ul^+l^-$ is induced by the $\gamma^*$ and $Z$ penguin
diagrams and  $WW$ box diagrams, the main contribution
again coming  form the intermediate $d$ and $s$ quarks. In contrast to $c\to 
u\gamma$, the $c\to ul^+l^-$ transition is not strongly GIM suppressed at one 
loop  \cite{vll}
\begin{eqnarray}
\label{ll}
{\cal L}_{SD}^{c\to ul^+l^-}= \frac{G_F}{{\sqrt 2}} {e^2 \over 
8\pi^2\sin^2\theta_W}c_7
\bar u \gamma^{\mu}(1-\gamma_5)c~\bar l \gamma_{\mu}l~,\qquad  c_7=-0.065
 \end{eqnarray}
giving the rate   $\Gamma(c\to
ul^+l^-)/\Gamma(D^0)\sim 2.9\times 10^{-9}$. QCD corrections  to Eq. \ref{ll} 
have not been determined yet, but they are not expected to be
sizable. 

\section{Long distance contributions}

In addition to a quark transition at short distances, a given hadron decay is 
contaminated by LD contributions. We examine different  {\bf meson decays} and  
briefly comment on the baryon decays in the next section. The decays of 
interest have the flavour content $c\bar q\to u\bar q\gamma$ and $c\bar q\to 
u\bar q\gamma^*\to u\bar q l^+l^-$, where  $u\bar q$ forms a pseudoscalar or a 
vector meson in the case of $l^+l^-$ and a vector meson in the case of a real 
photon $\gamma$ (in this case pseudoscalar is forbidden by gauge invariance). We 
will choose the flavour of $q=u,~d,~s,~c,~b$ so that LD contributions will be 
as small as possible. The emission of the final photon (real or virtual) can 
proceed {\it resonantly} via the $\rho$, $\omega$, $\phi$ exchange (photon 
emission from the light quarks) giving the resonant shape in the $m_{ll}$ 
spectrum or {\it nonresonantly}. 

The most serious is the {\bf LD pole} contamination, which arises via the $W$ 
exchange in s-channel for the case of $q=d,~s,~ b$  (sketched in Fig. 1 b) and 
via $W$ exchange in t-channel for the case of $q=u, ~c$ when in addition a 
photon is emitted from the initial $c\bar q$ or final $u\bar q$ meson.   The 
$W$ exchange  and corresponding QCD corrections are incorporated in the 
Lagrangian
\begin{equation}
\label{LD}
{\cal L}_{LD}=-{G_F\over \sqrt{2}}V^*_{cq}V_{uq}[a_1\bar 
q\gamma^{\mu}(1-\gamma_5)c~\bar u \gamma_{\mu}(1-\gamma_5)q+a_2\bar u 
\gamma^{\mu}(1-\gamma_5)c~\bar q\gamma_{\mu}(1-\gamma_5)q]~.
\end{equation}  
The magnitude of pole contribution depends on the flavour $q=d,~s,~b$ mainly 
through  $V^*_{cq}V_{uq}$ (see Fig. 1b) giving 
$|V^*_{cd}V_{ud}|=|V^*_{cs}V_{us}|=0.22$ for $q=s,~d$ and $|V^*_{cb}V_{ub}|\sim 
0.0002$ for $q=b$. Taking $q=b$ therefore minimizes the LD effects and the 
least contaminated meson decay to study $c\to u\gamma$ is $B_c\to B_u\gamma$ 
decay \cite{bc}. In $D$ meson decays  ($q=d,~s,~ u$) on the other hand, the  
pole contribution turns out to be several orders of magnitude larger then the 
SD contribution \cite{vll,FPS,FS}. 
Note that for $q=c$ there is no pole contribution, but $\eta_c$ can decay 
electromagnetically and the weak decay of interest is completely overshadowed in 
the experiment. 

The second is the {\bf LD vector meson dominance (VMD)} contribution sketched in 
Fig. 1b. Here the transition $c\to u\bar dd(\bar ss)$ is induced by the second 
term in Eq. \ref{LD},   the intermediate $\bar dd$ and $\bar ss$  hadronize in 
the  vector mesons $\rho^0$, $\omega$ and $\phi$, which finally  transverse to a 
photon. 
This contribution turns out to be proportional to the flavour $SU(3)$ breaking, 
so it is smaller than the  pole contribution.  Note that VMD contribution is 
present in any decay of interest and it  does not depend strongly on the 
flavour of the quark $q$, since $\bar q$ is merely a spectator (see Fig. 1b).

\section{Results}

Considering the LD pollution, the $B_c\to B_u^*\gamma$ decay presents the most 
suitable probe for $c\to u\gamma$ transition. The decay has been studied using 
the  factorization approximation and Isgur-Scora-Grinstein-Wise nonrelativistic 
constituent quark model \cite{bc}, which is considered to be reliable for a 
state $B_c$ composed of two heavy quarks.  The predicted SD and LD parts of the 
branching ratio, shown in Table 1, are of comparable size $\sim 10^{-8}$, which 
in principle allows to probe the $c\to u\gamma$ transition in $B_c\to 
B_u^*\gamma$ decay. Experimental detection of $B_c\to B_u^*\gamma$ decay at the 
branching ratio well above $10^{-8}$ would clearly indicate a signal for new 
physics.  The measurement of this decay would probe different scenarios of  
physics beyond the standard model: the non-minimal supersymmetric model 
\cite{BGM} and the standard model with four generations \cite{BHLP}, for 
example, predict  $Br(c\to u\gamma)$ up to $10^{-5}$, which would enhance 
$Br(B_c\to B_u^*\gamma)$ up to $10^{-6}$. Such effects could be detected at LHC 
\cite{LHC}, where $2.1\times 10^{8}$ mesons $B_c$ with $p_T(B_c)>20 ~GeV$ will 
be produced at integrated luminosity $100~fb^{-1}$.      

Similarly, $c\to ul^+l^-$ will be least contaminated by LD effects in $B_c\to 
B_ul^+l^-$ decay, but here one has to go to  the $m_{ll}$ region bellow the 
resonances $\rho$, $\omega$ and $\phi$. 

The $D$ meson decays $D\to V\gamma$ \cite{FPS,FS} and $D\to Vl^+l^-$ \cite{vll} 
($V$ is a light vector meson) have been studied using the factorization 
approximation and the model, which combines the heavy quark effective theory 
and chiral perturbation theory. The predicted  SD and LD branching ratios for 
Cabibbo suppressed FCNC decays are presented Table 1 (for completeness also non 
FCNC decays are presented). They are completely dominated by LD effects and even 
huge non-standard effects mentioned above would hardly be visible here. These 
decays can be therefore used as a controlled laboratory for LD effects, which 
present an important background to extract $b\to s\gamma(l^+l^-)$ in $B$ 
decays. The SD (dot-dashed) and LD (solid)  differential branching ratios  
$d\Gamma(D^0\to \rho^0\mu^+\mu^-)/dm_{\mu\mu}^2\Gamma_D$ are plotted in Fig. 2 
and one expects that above the resonance $\phi$ the LD contribution is reduced. 
In $D\to \pi l^+l^-$ decays the $m_{ll}^{max}=m_D-m_{\pi}$ is as high as 
possible and the region above resonance $\phi$, nonexistent in other decays, is 
dominated by short distance $c\to ul^+l^-$ process \cite{pll}. The decays $D\to 
\pi l^+l^-$ at $m_{ll}>1.25~GeV$ therefore present another possibility where 
$c\to ul^+l^-$ might be investigated in the future \cite{pll}.  

Having exhausted the meson decays, let us comment on the baryon decays. The VMD 
contribution is present in every decay of interest, while potentially more 
dangerous W exchange  pole contribution is absent in decays of the baryons, 
where all the valence quarks have the same charge. The least exotic decay to 
look for $c\to u\gamma$, which is not to contaminated by LD effects, is 
$\Sigma_c^{++}\to \Delta^{++}\gamma$ ($cuu\to uuu\gamma$), but the strong decay 
channel $\Sigma_c^{++}\to \Lambda_c^+\pi^+$ completely overshadows the weak 
decays. We are left only with the more exotic decays $\Xi_{cc}^{++}\to 
\Sigma_c^{++}\gamma$  and $\Omega_{ccc}^{++}\to \Xi_{cc}^+\gamma$ as probes for 
$c\to u\gamma$. 
   
\section{Conclusion}

In general it is difficult to observe the $c\to u$ transition. The $c\to u\gamma$ transition can be probed in $B_c\to B_u^*\gamma$ decay 
and the measurement of its branching ratio well above $10^{-8}$ would signal 
new physics. The $c\to ul^+l^-$ transition may be probed at high $m_{ll}$ in 
$D\to \pi l^+l^-$ decay or at low $m_{ll}$ in $B_c\to B_ul^+l^-$ decay.  

\begin{table}[t]
\begin{center}
\begin{tabular}{|c|c|c||c|c|}
\hline
FCNC decay & $Br_{SD}$ & $Br_{LD}$ & non FCNC decay &  $Br_{LD}$\\
\hline
$B_c\to B_u^*\gamma$ &$4.7\times 10^{-9} $ [ \cite{bc}]&$7.5({+7.7\atop 
-4.3})\times 10^{-9}$ [ \cite{bc}]& &\\
\hline 
$D^0\to \rho^0\gamma$ & &$[0.1-1.0]\times 10^{-5}$ [ \cite{FPS}]&$D^0\to \bar 
K^{*0}\gamma$ &$[0.6-3.6]\times 10^{-4}$ [ \cite{FPS}]\\
$D^0\to \omega\gamma$ &of&$[0.1-0.9]\times 10^{-5}$ [ \cite{FPS}]&$D_s^+\to 
\rho^+\gamma$ &$[2.0-8.0]\times 10^{-4}$ [ \cite{FPS}]\\
$D^0\to \phi\gamma$ &order &$[0.4-1.9]\times 10^{-5}$ [ \cite{FPS}]&$D^+\to 
K^{*+}\gamma$ \c&$[0.3-4.4]\times 10^{-6}$ [ \cite{FPS}]\\
$D^+\to \rho^+\gamma$ &$10^{-9}-10^{-8}$&$[0.4-6.3]\times 10^{-5}$ [ 
\cite{FPS}]&$D^0\to K^{*0}\gamma$ &$[0.3-2.0]\times 10^{-6}$ [ \cite{FPS}]\\
$D_s^+\to K^{*+}\gamma$ &&$[1.2-5.1]\times 10^{-5}$ [ \cite{FPS}]&&\\
\hline
$D^0\to \rho^{0}\mu^+\mu^-$ &&$[3.5-4.7]\times 10^{-7}$ [ \cite{vll}]&$D^0\to 
\bar K^{*0}\mu^+\mu^-$ &$[1.6-1.9]\times 10^{-6}$ [ \cite{vll}]\\
$D^0\to\omega\mu^+\mu^-$&of &$[3.3-4.5]\times 10^{-7}$ [ \cite{vll}]&$D_s^+\to 
\rho^{+}\mu^+\mu^-$&$[3.0-3.3]\times 10^{-5}$ [ \cite{vll}]\\
$D^0\to\phi\mu^+\mu^-$&order&$[6.5-9.0]\times 10^{-8}$ [ \cite{vll}]&$D^+\to 
K^{*+}\mu^+\mu^-$&$[3.1-3.7]\times10^{-8}$ [ \cite{vll}]\\
$D^+\to \rho^+\mu^+\mu^-$&$10^{-10}-10^{-9}$&$[1.5-1.8]\times 10^{-6}$ [ 
\cite{vll}]&$D^0\to K^{*0}\mu^+\mu^-$&$[4.4-5.1]\times 10^{-9}$ [ \cite{vll}]\\
$D_s^+\to K^{*+}\mu^+\mu^-$& [ \cite{vll}]&$[5.0-7.0]\times 10^{-7}$ [ 
\cite{vll}]&&\\
\hline
\end{tabular}
\end{center}
\caption{The predicted short distance (SD) and long distance (LD) parts of branching ratio for FCNC decays are presented on the left; non FCNC decays are given on the right. The 
experimental upper bounds for $Br(D\to V\gamma)~~^{12}$ 
are at the level 
$10^{-4}$, so this decays might be detected soon. For $Br(D\to Vl^+l^-)~~^{1,13}$ 
they 
 are at the level $10^{-3}-10^{-4}$, but there is no existing 
bound for $D_s^+\to \rho^+l^+l^-$, which has the best chances of detection. The 
error bars in $B_c$ channel arise form the error in $\Gamma(V^0\to e^+e^-)$ data 
used in determination of VMD contribution, which is proportional to flavour 
SU(3) breaking $^2$
. In $D$ decays the errors are due to the uncertainty of the 
parameters in the effective model. }
\end{table}

\begin{figure}
 \begin{center}
 \leavevmode\epsfbox{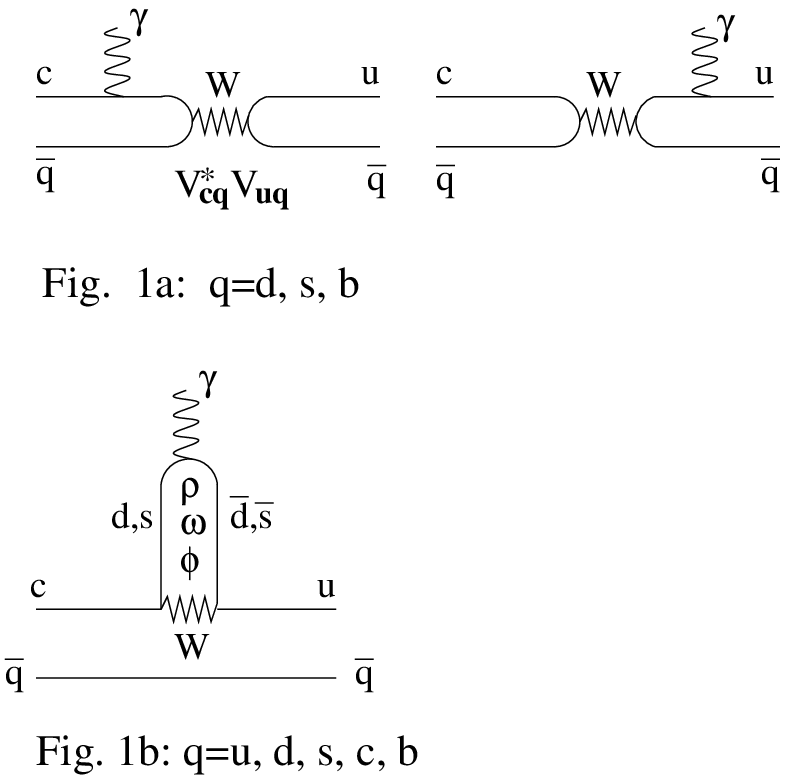} 
 \caption{Skeleton diagrams of long distance a) {\it pole} contribution (photon 
is emitted from the meson as a whole) and b) {\it VMD} contribution. } 
 \end{center}
  \end{figure}

\begin{figure}
 \begin{center}
\epsfxsize=7.5cm
 \epsfysize=6.5cm
 \epsffile[0 0 490 380]{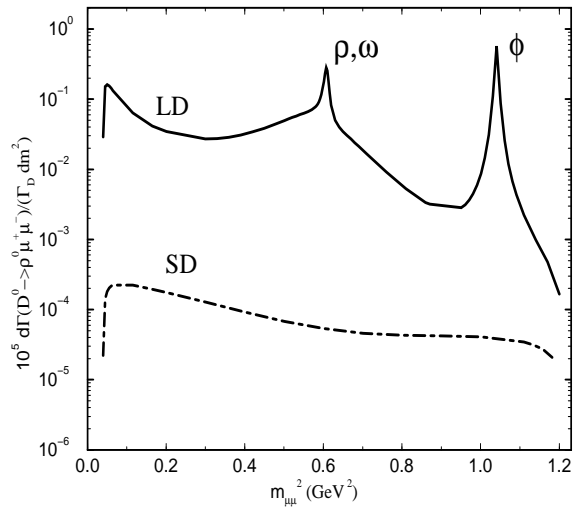}  
 \caption{The SD (dot-dashed) and LD (solid) contribtions to 
$(1 /\Gamma_D ) d \Gamma (D^0 \to \rho^{0} \mu^+ \mu^-)/ d m_{\mu\mu}^2$ 
as a function of $m_{\mu\mu}^2$ as predicted in $^6$. } 
 \end{center}
  \end{figure}

\end{document}